\documentclass[12pt]{iopart}
\usepackage[dvips]{graphicx}

\begin{document}
\title[2D versus 3D Freezing of a LJ Fluid in a Slit Pore]{2D versus 3D Freezing of a Lennard-Jones Fluid in a Slit Pore: A Molecular Dynamics Study}
\author{N Gribova$^1,^2$, H Maleki$^3$, A Arnold$^4$, C Holm$^1$, T Schilling$^3$}
\address{$^1$ Institute for Computational Physics, University of Stuttgart,
Pfaffenwaldring 27, D-70569 Stuttgart, Germany}
\address{$^2$ Institute for High Pressure Physics, Russian Academy of Sciences, Troitsk 142190,
Moscow Region, Russia}
\address{$^3$ Institut f\"ur Physik, Johannes Gutenberg-Universit\"at,
  Staudinger Weg 7, D-55099 Mainz, Germany}
\address{$^4$ Fraunhofer SCAI, Schloss Birlinghoven, D-53754 Sankt Augustin, Germany}
\ead{gribova@icp.uni-stuttgart.de}


\begin{abstract}
We present a computer simulation study of a (6,12)-Lennard-Jones fluid
confined to a slit pore, formed by two uniform planes. These interact via a (3,9)-Lennard-Jones potential with  the
fluid particles. When the fluid approaches the liquid-to-solid transition we first observe
layering parallel to the walls. In order to investigate the nature of the freezing transition we performed a detailed analysis of the bond-orientational order parameter in the layers. We found no signs of hexatic order which would indicate a melting scenario of the Kosterlitz-Thouless type. An analysis of the mean-square displacement shows that the particles can easily move between the layers, making the crystallization a 3d-like process.
This is consistent with the fact that we observe a considerable hysteresis
in the heating-freezing curves, showing that the crystallization transition proceeds as an
activated process.
\end{abstract}

\pacs{}
\submitto{\JPCM}
\maketitle

\section{Introduction}
Understanding the structure and dynamics of confined fluids is
important for processes such as wetting, coating, and nucleation. The
properties
of a fluid confined in a pore differ significantly from the bulk fluid due to
finite size effects, surface forces and reduced dimensionality. In this
work we report a study of one of the simplest models that is still
capable of reproducing the thermodynamic behavior of classical fluids,
the Lennard-Jones (LJ) system. The LJ
potential is an important model for exploring the behavior of
simple fluids and has been used to study homogeneous vapor-liquid,
liquid-liquid and liquid-solid equilibria, melting and freezing. It has
also been used as a reference fluid for complex systems like colloidal
and polymeric systems.

The vapor-to-liquid transition in confined systems has been studied
intensively, and it is well understood (see \cite{Gelb_review1999} and
references therein).
In this article we will discuss the liquid-to-solid transition in a
slit pore and the process of the development of the solid phase.
In the liquid phase, confinement to a slit induces layering at the walls.
One could imagine this effect to facilitate crystallization. And indeed it
is known that depending on the strength of the particle-wall interaction
different scenarios of freezing exist \cite{Miyahara1997,Alba2006}.
If the walls are strongly attractive, crystallization starts from the walls and
at a temperature higher than without confinement. If, however,
the walls are strongly repulsive,  crystallization starts from  the bulk at a temperature lower than
without confinement. A well-distinguished layer of
particles at the wall can also, to some extent, be treated as a 2d system.
This raises the question, whether freezing of such a layer proceeds via the
Kosterlitz-Thouless-Halperin-Nelson-Young (KTHNY) mechanism
\cite{KT73,halperin78,nelson79,young79}, meaning that the
liquid turns into a crystal going through a hexatic
phase \cite{strandburg}. This question has been studied for rather narrow
pores (up to 7.5 diameters of a fluid particle).
It was found that a hexatic phase exists between liquid and crystal
only in the contact layers at the walls \cite{radhakrishnan2002JCP}.
As a consequence, in a pore that can accommodate only a single layer
the transition is of the KTHNY type. However, with increasing  width the
behavior changes
to a first order transition \cite{radhakrishnan2002}.

Here we investigate an attractive pore that is significantly wider, namely 20 diameters of
a fluid particle. Studying the bond-orientational order parameter within the layers we observe
no sign of a hexatic phase. An analysis of the mean-square displacement shows that the particle diffuse between the layers. Hence, the crystallization proceeds as a 3d process, as is also suggested by the noticable hysteresis loop in the heating-freezing curve.

The article is structured as follows. First we describe our simulation method. In section \ref{sec:results}. we present the results, followed by a discussion, and in section \ref{sec:conclusions}. we conclude with a summary of the presented results.

\section{Simulation method}
We performed molecular dynamics (MD) simulations in the isothermal
ensemble (NVT), i.e. the number
of particles N, the volume V and the temperature T were fixed.
The system consists
of 8000 particles confined between two structureless walls. The particles
interact via the LJ-potential
\begin{equation}
 u(r)=4 \epsilon \left[ \left( \frac{\sigma}{r}\right)^{12}-\left(\frac{\sigma}{r}\right)^{6}\right].
\end{equation}
The interaction between walls and particles is characterized by a LJ-potential
integrated over semi-space:
\begin{equation}
 u_w(r)=4\epsilon\left[ \left( \frac{\sigma}{r}\right)^{9}-\left(\frac{\sigma}{r}\right)^{3}\right].
\end{equation}
The particle-particle interaction was cut off at a distance $r_c=2.5 \sigma$ and the
wall-particle interaction at a distance  $r_c=4.0\sigma$ (the wall-particle
potential is wider and deeper then the particle-particle potential). Using
for the wall-particles interaction a Steelle potential \cite{steelle}
or just a (4,10)-LJ potential
does not influence the results qualitatively.
For the following we will use $\epsilon$
as the unit of energy, $\sigma$ as the
unit of length and $\tau=\sqrt{1 \cdot \sigma^{2}/\epsilon}$ as unit of time (i.e. use the
particle mass as the unit of mass); consequently, temperatures are
given in multiples of $\epsilon/k_{\mathrm{B}} T$.
The simulations were performed in a cubic box with periodic boundary
conditions in the x and y direction. The walls were positioned at
$z=0$ and $z=L_z=20\sigma$. The size of the simulation box was $L_x=L_y=L_z=20
\sigma$. We used standard Nos{\'e}-Hoover and Langevin
thermostats to keep the temperature constant \cite{rapaport, allen,
  frenkel}.
For the Langevin thermostat, the friction was always chosen as
$\Gamma=m\tau^{-1}=1$~\cite{allen}. 
For the Nos{\'e}-Hoover thermostat, we set the effective mass $M_s=0.5$. 
We simulated a cooling curve starting out from a random configuration at
$T=3.0$ and a melting curve starting out from a face-centered-cubic
configuration at $T=1.2$.
Far away from the transition, the temperature was changed by $\Delta T=0.1$
from one simulation run to the next, while close to transition
we used a smaller increment/decrement, $\Delta T=0.01$.

The simulations were performed with a timestep of
$\Delta t = 0.005$ and let run for $1.0\times10^{6}$ MD steps
for equilibration and for $2.5\times10^{5}$ for sampling. In order to compute
the mean square displacement of the particles, a smaller timestep
$\Delta t = 0.002$ was chosen, and
to avoid influence of thermostat on the dynamics of the system
we switched to the NVE ensemble after equilibration (keeping the total
energy of the system $E$ constant). We monitored the
temperature, which fluctuated around a mean value practically
equal to the temperature T in the NVT ensemble.
Pressures were obtained from the virial expansion \cite{hansen} omitting
corrections for the cut-off in the potential.
For parts of our simulations the software package ESPResSo version 2.04s was used \cite{limbach06a}.

\section{Results and Discussion \label{sec:results}}

\begin{figure}
\begin{center}
\includegraphics[width=9cm]{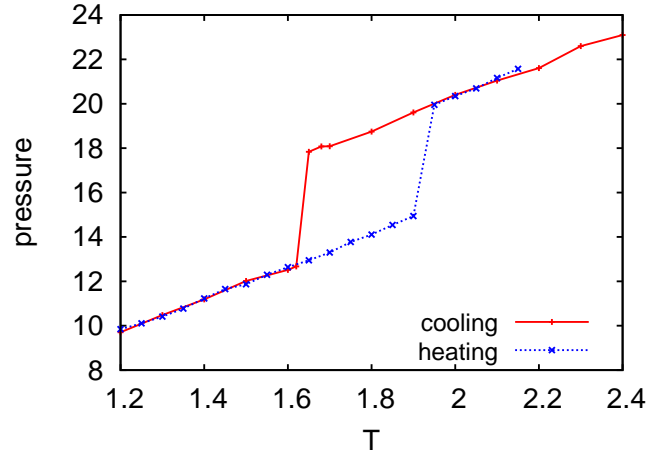}
\caption{\label{fig:pressure}Pressure-Temperature curves for heating and
  cooling of a LJ-fluid confined between two uniform LJ-walls. The outer borders
borders of hysteresis are shown.}
\end{center}
\end {figure}

In Figure~\ref{fig:pressure}, the pressure-temperature curves for heating and
cooling are shown. There is a considerable hysteresis, which indicates that the
system has to overcome a  free energy barrier when transforming
from one phase to the other.
Several runs were performed both for heating and for cooling. The cooling
lines coincide, whereas the temperature at which the melting process starts
fluctuates. In the Figure~\ref{fig:pressure} we show the outer borders of the
hysteresis region.

\begin{figure}
\begin{center}
\includegraphics[width=7cm]{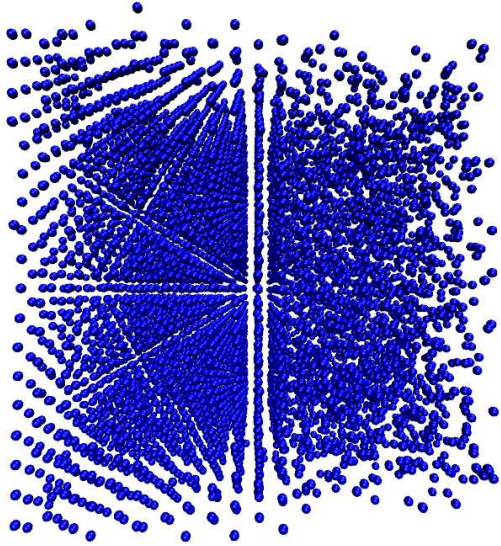}
\caption{\label{fig:snapshot} Snapshot of the system in the early stage of crystallization  at T=1.60.
In this specific example it crystallizes from the left wall.
Walls are not represented, but are to the right and to the left of the box. The part on
the left is already a crystal while the right side is still disordered.}
\end{center}
\end {figure}

As our system has attractive walls, crystallization should
start from the walls \cite{Miyahara1997}. This can be clearly seen in the snapshot
(Figure~\ref{fig:snapshot}) that was taken 90 MD steps after equilibration had started.
Only another 90 MD steps later the system completely crystallized. One can
also see that no crystallization process has started at the right wall yet,
demonstrating that this event is an activated process.

As it was shown in \cite{Miyahara1997} the width of hysteresis depends on
the distance between the layers. If the distance differs considerably from the
lattice constant of an ideal  LJ crystal ($0.916$), then the hysteresis will
be more pronounced. For our system the distance is $0.87$ in the bulk, and
correspondingly the hysteresis is quite wide.

\begin{figure}
\begin{center}
\includegraphics[width=2.3 in, angle=270]{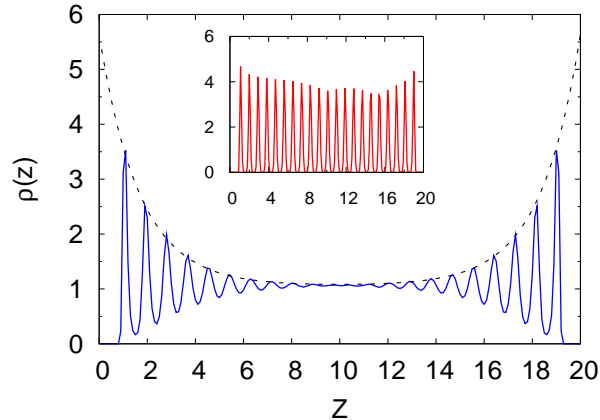}
\caption{\label{fig:profile} Density profile $\varrho(z)$ of the liquid
phase at T=1.70. The heights of the peaks are fitted by an exponential
function. The inset shows $\varrho(z)$ for the solid phase at T=1.60.}
\end{center}
\end {figure}

\begin{figure}[htb]
\begin{center}
\includegraphics[width=10cm]{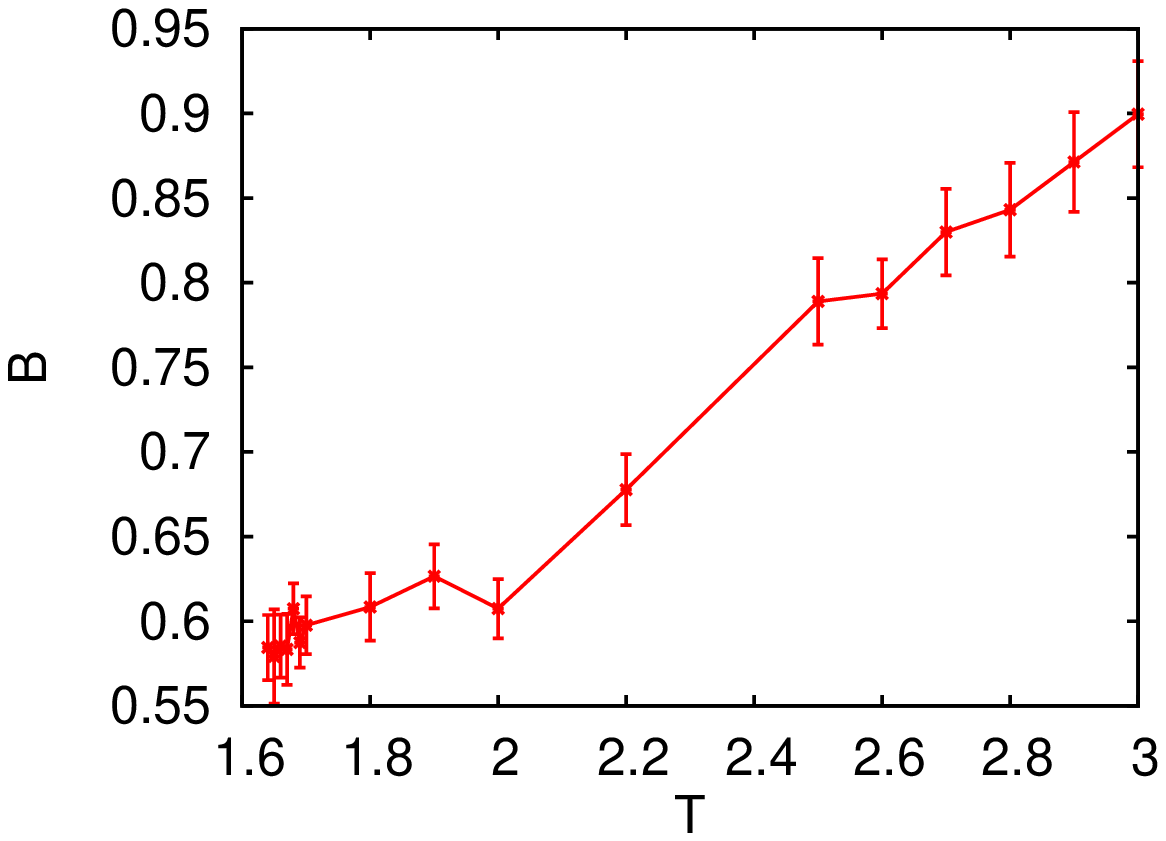}
\caption{\label{fig:fitrdf} Coefficient $B$ characterizing the exponential
decay
of density profile. In the hysteresis region $(T<=2.0)$ it almost not changing,
i.~e.~the structure of the density profile stays the same.}
\end{center}
\end{figure}

In order to investigate the phase transformation process, we now turn to the
effects the walls have on the structure of the fluid.
Figure~\ref{fig:profile} shows number density profiles $\varrho(z)$
for $L_{z}=20.0\sigma$ in the liquid and the solid phase.
In the liquid phase, the maxima of the peaks follow an exponential law
$A \left[\exp(-B x)+ \exp (-(L_z-x) B)\right]+\rho_{\mathrm {mid}}$, where
$\rho_{\mathrm {mid}}$ is the density in the middle of the box.
Figure~\ref{fig:fitrdf} shows the behavior of the coefficient B with temperature.
It can be seen that the values of B decrease more or less linearly at first, i.~e.~the number of layers increases
and they become more pronounced.
 As soon as we enter the regime of the hysteresis at $T=2.0$, $B$ becomes
almost constant (within the error of the simulations). This shows that the
structure of the density profile does not change, no new layers appear and
the system is trapped in the undercooled state.

As the liquid forms layers, one could assume that the transformation
proceeds inside the layers via a KTHNY transition. In order
to test this assumption, we now turn to the structure within the layers:
To characterize the transitional order in one layer in 2D, we calculate the
pair correlation function:
\begin{equation}
 g(r)=\varrho^{-2}\langle\sum _{i,i\neq j}\delta(\textbf{r}_{i})\delta(\textbf{r}_{j}-\textbf{r})\rangle
\end{equation}
where $\varrho$ is the number density of particles in each layer.
\begin{figure}
\begin{center}
\includegraphics[width=2.3 in, angle=270]{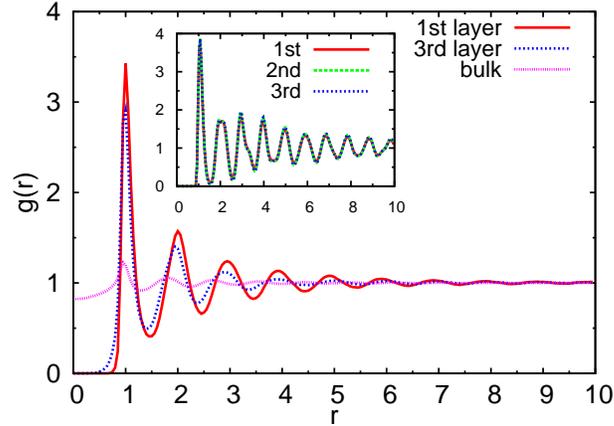}
\caption{\label{fig:rdf_layer} 2D pair correlation function in
  the first and third layer and the bulk part of the system in the
  the liquid state at $T=1.65$. The inset shows $g(r)$ for the first
three layers at the wall
for the solid state at $T=1.60$}
\end{center}
\end{figure}
In Figure~\ref{fig:rdf_layer} the 2D radial distribution functions for the first and third layer (seen from the wall), the bulk part
of the liquid and the first three layers of the solid phase are shown.
The structure within the layers of the liquid becomes less pronounced as we move further
away from the walls and is barely visible in the center of the box.

Next we consider the bond-orientational order \cite{strandburg}: we define the local
bond-orientational order parameter of
particle $j$ in layer $m$ at a position $\textbf{x}_{j}$ as
\begin{equation}
 \psi^m_{6}(\textbf{x}_{j})=\frac{1}{N_{j}}\sum_{k=1}^{N_{j}}e^{i6\theta_{jk}}
\end{equation}
where $N_{j}$ is the number of neighbors of particle $j$ within layer $m$, 
the sum is over the neighbors $k$ of $j$ within $m$, 
and $\theta_\textit{jk}$ is the angle between
an  arbitrary fixed axis and the line connecting particles j and k. The order
of the $m$-th layer $\Psi^m_6$ is defined as the average over
$\psi^m_{6}(\textbf{x}_{j})$ for all $N_m$ particles within the layer
\begin{equation}
 \Psi^m_6=\frac{1}{N_m}|\sum^{N_m}_{j=1}\psi^m_{6}(\textbf{x}_{j})| \quad .
\end{equation}

\begin{figure}
\begin{center}
\includegraphics[width=2.3 in, angle=270]{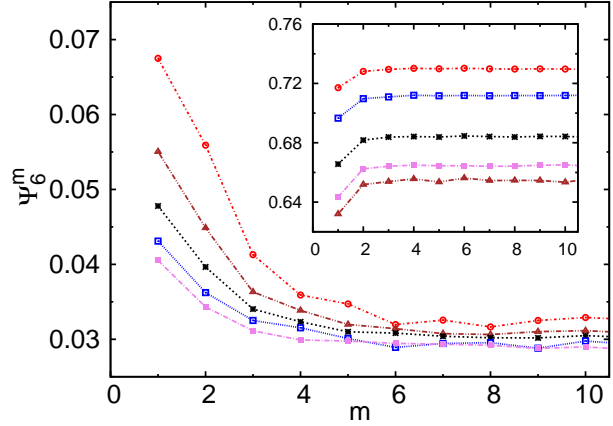}
\caption{2D bond-orientation order parameter $\Psi^m_6$ for liquid states depending on the
layer $m$ for temperatures 1.70, 1.80, 2.00, 2.20, 2.40 (from top to bottom).
The inset shows $\Psi^m_6$ for solid states
for temperatures 0.80, 1.00, 1.30, 1.50, 1.60 (from top to bottom).} \label{fig7}
\end{center}
\end {figure}

Figure~\ref{fig7} shows $\Psi^m_6$ for various temperatures. When approaching
the transition, the bond-orientational order close to the wall increases.

\begin{figure}
\begin{center}
\includegraphics[width=2.3 in, angle=270]{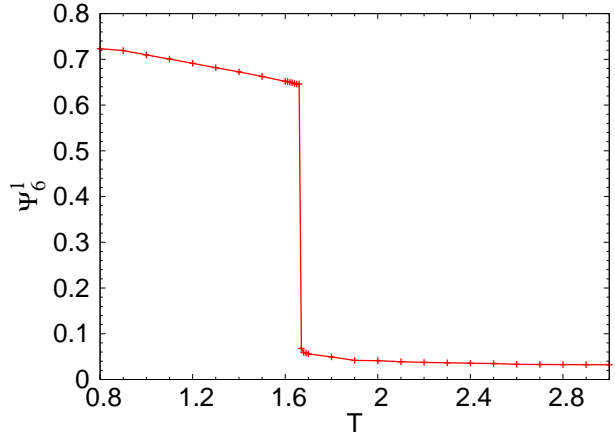}
\caption{2D bond-orientation order parameter $\Psi^1_6$ of the first layer
  from the wall as a function of the temperature.} \label{fig8}
\end{center}
\end {figure}

The temperature dependence of $\Psi^1_6$ for the first layer of particles
at the wall is shown in Figure~\ref{fig8}. It clearly ``jumps'' i.~e.~is
discontinuous at the transition.

If the crystallization proceeded purely within the two-dimensional layers,
one would observe a hexatic phase, which is characterized by a
power-law decay of the correlation of the bond-orientational order

\begin{equation}
g_{6}(r)=\langle\psi^{*}_{6}(\textbf{x}^{'})\psi_{6}(\textbf{x}^{'}-\textbf{x})\rangle
\quad ,
\end{equation}
where the average is taken over all particles within a layer whose positions $\textbf{x}$
are a distance $r$ apart.

\begin{figure}
\begin{center}
\includegraphics[width=2.3 in, angle=270]{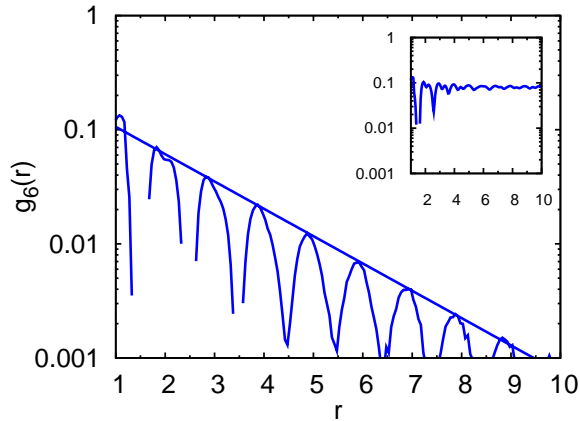}
\caption{Bond-orientational correlation function for the liquid state
just before freezing at
  $T=1.67$ for the  layer closest to the wall. The solid line is the result of an exponential fit.
There is no signature of a hexatic phase. The inset shows
$g_6$ for the crystallin state at the next available lower
temperature  $T=1.66$, right after the transition. } \label{fig9}
\end{center}
\end {figure}

Figure~\ref{fig9}  shows $g_{6}(r)$
for the first layer at $T=1.67$ and $T=1.66$. The system jumps from the
2d-liquid phase into the 2d-solid without visiting a hexatic phase
first. Hence the crystallization process is not of the KTHNY-kind.

\begin{figure}[htb]
\begin{center}
\includegraphics[width=10cm]{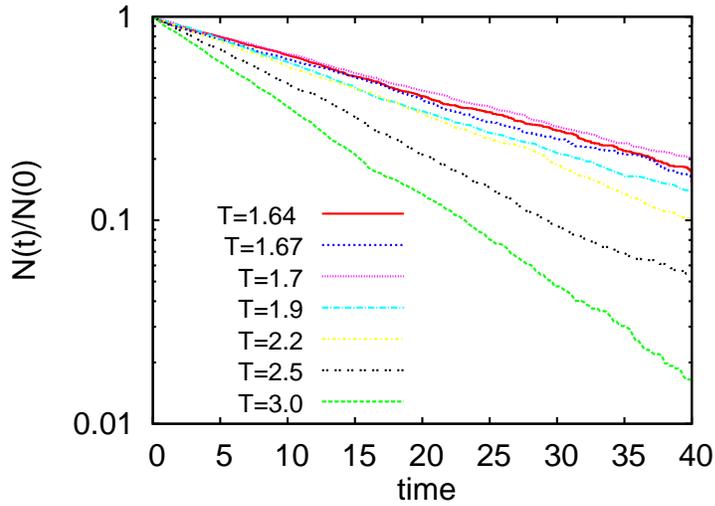}
\caption{\label{fig:lifetime}  Ratio of particles in the layer closest to the wall that stayed
there from time $0$ ($N(0)$) until time $t$ ($N(t)$) at different
temperatures above
the phase transition.}
\end{center}
\end{figure}

\begin{figure}[htb]
\begin{center}
\includegraphics[width=10cm]{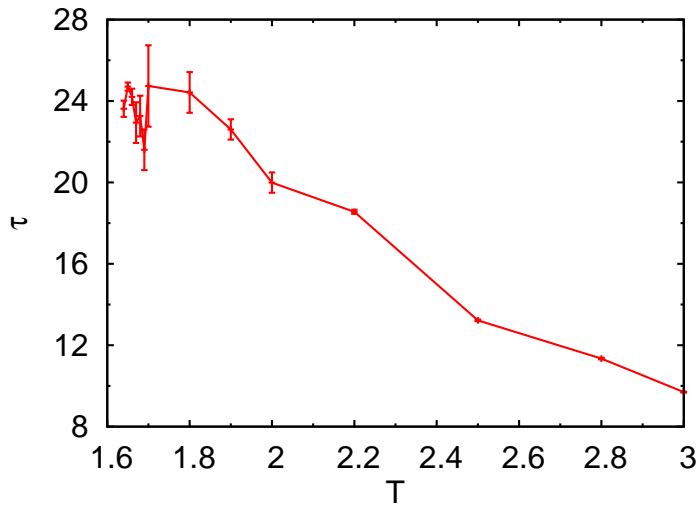}
\caption{\label{fig:lifefit} Average lifetime of particles in the outmost layer as a function of temperature.
In the hysteresis region the lifetime does not change significantly.}
\end{center}
\end{figure}

To find out why the crystallization process is 3d-like despite the layering,
we now consider the particles' dynamics. One of the obvious characterictics is to
estimate how long particles on average  stay in the layer closest to the
wall. The easiest way
to estimate this is to calculate how many
particles of those which were in the layer at time 0 remained there at the time $t$.
From Figure~\ref{fig:lifetime} we can see that the ratio of particles that
remain in the layer decreases exponentially
with time. Fitting it  with $\exp(-t/\tau)$ we obtain the average
lifetime $\tau$ of a particle in
the layer (Figure~\ref{fig:lifefit}). It increases linearly with the decrease of
temperature and is then fluctuating around the mean value in the hysteresis
region. As we observed already for the density, the behavior of the system in the hysteresis
region does not change much during cooling.

\begin{figure}[htb]
\begin{center}
\includegraphics[width=10cm]{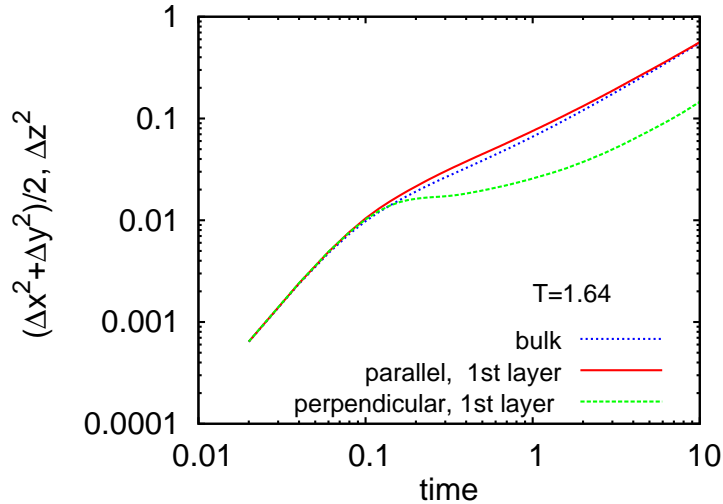}
\caption{\label{fig:msd164} Mean square displacement for $T=1.64$. The MSD parallel to the
wall is almost identical in the layer and in the bulk with the particles in the layer even
being slightly faster. The MSD perpendicular to the wall show a clear trapping effect.}
\end{center}
\end{figure}

To characterize the mobility of the particles we calculated the mean
square displacement (MSD).
As the system forms layers, we calculate the MSD parallel and perpendicular to
the wall separately. Looking at the plane parallel to the wall
(Figure~\ref{fig:msd164})  while approaching crystallization, we observe that
 the particles in the layer closest to the wall are a little faster than the particles
in the bulk, despite the fact that the crystallization typically starts from here.
We take this as another hint that the crystallization proceeds as a 
3d-process, and does not first start within the layer closest to the wall.
The behavior of the particles does not change significantly on approach of
the crystallization as they enter the metastable region.

The mean square displacement
measured perpendicular to the wall (Figure~\ref{fig:msd164}) shows that after the ballistic regime
for a while particles are trapped in the layer and then start leaving it. It is not meaningful to calculate the diffusion coefficient in our system, because the particles
do not stay long enough in a layer for the MSD to enter the linear regime.


\section{Conclusions \label{sec:conclusions}}

We reported on a molecular dynamics study of the liquid-to-solid
transformation of a LJ fluid in a wide slit pore. Although the
confinement induces layering in the liquid phase close to the walls,
we do not find a successive, layerwise
crystallization. Crystallization is still a 3d process, and, in
particular, no hexatic phase was observed in the layers closest to the
wall, excluding the possibility of a 2D KTHNY-like crystallization
within the layers; in fact, the mobility of particles in the layers is
higher than their mobility in the bulk. Nevertheless, we find that
crystallization in the system practically always starts from the
walls, i.~e., the walls facilitate crystallization. And although
crystallization is an activated process similar to 3d crystallization,
we observe a smaller hysteresis, indicating a reduced
nucleation barrier as compared to bulk crystallization.

Alltogether, our simulations suggest that the nucleation of
the LJ fluid close to a planar wall does not significantly differ from
the nucleation in the bulk, although with a smaller nucleation
barrier. This can however be easily understood as an effect of the
strongly increased density in the layers close to the confinement.

\ack
NG thanks Dr. Marcello Sega for fruitful discussions. The authors thank the DFG
for financial support through the SPP1296 and an Emmy Noether award to TS.

\bibliographystyle{unsrt}
\bibliography{paper}

\end{document}